# Combining Coarse-Grained Protein Models with Replica-Exchange All-Atom Molecular Dynamics

**Jacek Wabik, Sebastian Kmiecik, Dominik Gront, Maksim Kouza \* and Andrzej Koliński \***

Faculty of Chemistry, University of Warsaw, L. Pasteura 1, Warsaw 02-093, Poland;
E-Mails: jwabik@chem.uw.edu.pl (J.W.); sekmi@chem.uw.edu.pl (S.K.);
dgront@chem.uw.edu.pl (D.G.)

\* Authors to whom correspondence should be addressed; E-Mails: mkouza@chem.uw.edu.pl (M.K.); kolinski@chem.uw.edu.pl (A.K)

**Abstract:** We describe a combination of all-atom simulations with CABS, a well-established coarse-grained protein modeling tool, into a single multiscale protocol. The simulation method has been tested on the C-terminal beta hairpin of protein G, a model system of protein folding. After reconstructing atomistic details, conformations derived from the CABS simulation were subjected to replica-exchange molecular dynamics simulations with OPLS-AA and AMBER99sb force fields in explicit solvent. Such a combination accelerates system convergence several times in comparison with all-atom simulations starting from the extended chain conformation, demonstrated by the analysis of melting curves, the number of native-like conformations as a function of time and secondary structure propagation. The results strongly suggest that the proposed multiscale method could be an efficient and accurate tool for high-resolution studies of protein folding dynamics in larger systems.

## 1. Introduction

The computational modeling of proteins plays a major role in understanding various aspects of molecular biology. Protein structure prediction [1], folding [2] and unfolding [3], aggregation [4] and interactions with other biomolecules have been studied by computer simulations at different levels of resolution and timescales. For more efficient simulations coarse-grained (CG) models [2] are used—standalone or in combination with atomic-level molecular dynamics (MD) [5]. The CG models reduce the complexity of each amino acid representation by one pseudo atom [6] or a group of pseudo atoms [7–9]. The bead usually reflects the size of a specific fragment and sometimes its geometry (e.g., the CABS [8] or the UNRES model [7]).

In our work the CABS model [8] is used in which a protein chain is represented by CA, CB and side-chain pseudo atoms. This well-established method has been successfully applied to structure prediction [1] and studying the folding dynamics on long [10–13] and short [14] timescales.

However, the high performance of CG simulations comes at the cost of structural accuracy. To derive more detailed characterization of a protein folding pathway and conformational ensembles, all-atom force fields [15–18] have to be used. At present, however, folding simulations with such sophisticated models (especially in explicit solvent) remain computationally very demanding.

Numerous efforts have been made to develop methods (umbrella sampling [19], multi-canonical ensemble [20], replica-exchange (RE) [21–23]) which speed up the computation of thermodynamic quantities for a system investigated. Over the last two decades the RE method has become the most widely applied approach to enhance the sampling of biomolecules [24–29]. It is worth noting that replica-exchange molecular dynamics (REMD) is particularly helpful in implicit solvent simulations. Modeling in explicit solvent, though, remains challenging due to the system size and the consequent timescale limitations up to hundreds of nanoseconds. Such simulations are affected by two difficulties: (1) using a large number of water molecules requires that the chosen temperature intervals are very small. Many replicas are required even for small proteins to cover the range between the extreme temperatures. (2) The number of round trips (the walk of a replica in the temperature space from the lowest to the highest T and back) decreases with the system size or with a large number of replicas. As a consequence, simulations in explicit solvent require not only a large number of replicas, but also long simulation times to obtain sufficient statistics for the calculation of thermodynamic quantities.

Various strategies to accelerate the convergence of REMD simulations have been proposed such as coupling to a High-Temperature Structure Reservoir [30,31] and other [32]. Here we present a hybrid method in which CG Monte Carlo (MC) simulation is combined with all-atom modeling with the aim of circumventing the problem of long equilibration time in REMD simulations. It is implemented with reconstructed CG models as starting structures in the all-atom REMD simulation. A similar concept was used for isothermal trajectory simulations [33–37]. Here we extend this concept for REMD in explicit solvent. The effectiveness of the approach is demonstrated on a model peptide, the C-terminal β-hairpin from B1 domain of protein G (PDB code: 2GB1, residues 41 to 56), previously used in a number of studies, both computational [29,38–44] and experimental[44–50]. Our simulation results suggest that the proposed hybrid approach can reduce the time needed to reach equilibrium dramatically while retaining the accuracy of the traditional REMD technique.

## 2. Results and Discussion

*2.1. Analysis of REMD Simulation Convergence*

To characterize simulation convergence (see Methods for the simulation details), we analyzed how the number of replicas that were in a folded conformation changed over time. This method was applied earlier for investigating the equilibration time of the Trp-cage miniprotein by Paschek *et al.* [51,52] and for other peptides [53] and RNA [54]. Clearly, trajectories started from CABS-generated conformations converged faster than runs started from extended ones, in particular for the OPLS-AA force field. CABS-initiated simulations in this case equilibrated after the 10th nanosecond when the number of folded replicas reached a stationary level (Figure 1a). In the beginning (before 10 ns), the number of near-native conformations fluctuated from 10 to 15 and later it converged to the value of 11−12 with negligible deviations.

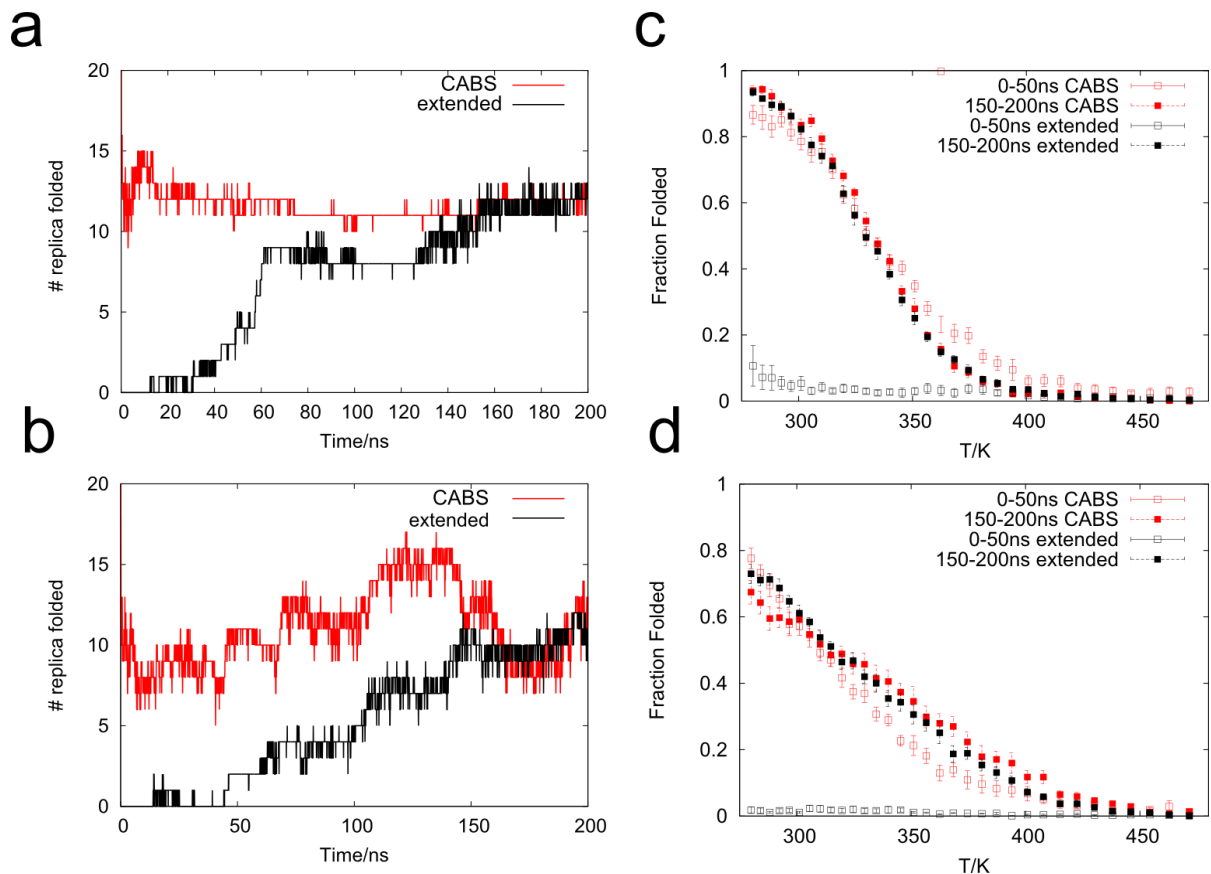

**Figure 1.** Number of folded replicas as a function of time [(**a**) OPLSAA force field; (**b**) Amber99sb force field.] and fraction of native-like conformations at corresponding temperatures [(**c**) OPLSAA force field; (**d**) Amber99sb force field] for two starting options: extended conformations (black) and CABS-generated conformations (red). Data from various time ranges were applied to indicate the time essential for reaching the equilibration state.

In contrast, the simulations that started from extended conformations needed much more time to reach the equilibrium. In the OPLS-AA force field, the number of folded replicas gradually increased from 0 to 8 at 60 ns. This value remained constant until 130 ns and then began to increase reaching a value of 11–12 at *ca.* 160 ns and it remained constant till the end of the simulation. The final number of folded replicas was consistent with simulations started from CABS conformations.

In the case of the Amber99sb force field (Figure 1b), the number of folded replicas was more variable during simulation time. This was also in accordance with other results [38] where it was evident that there were more transitions between native and non-native conformations for Amber99sb compared to the OPLS-AA force field. Thus it was more difficult to clearly define the moment when the simulation converged.

For further analysis we extracted two parts from the resulting trajectories: corresponding to 0–50 ns and 150–200 ns, respectively. Subsequently, for each part we computed a melting curve (Figure 1(c,d)) *i.e.*, a plot showing a fraction of folded conformations as a function of temperature. Each point in this plot represents an average value calculated from a single replica (*Y* axis) simulated at constant temperature (*X* axis). In general, the first 50 ns of REMD started from an extended conformation yielded nearly no native-like conformations. In the same period, starting from CABS-generated conformations we obtained 87% of folded conformations in a replica of 280 K for the OPLSAA force field. The fraction exponentially decreased at higher temperatures and finally reached nearly zero in replicas above *ca.* 400 K. In the final 50 ns at 280K we observed *ca.* 95% native-like conformations which agreed well with the experimental results of Munoz *et al.* [50] (ca. 80% at 280K). A comparable fraction of native-like conformations in the OPLS-AA force field was obtained by other authors [38, 55]. Similarly, after the first 50ns of simulation at 280K in the AMBER force field, the fraction of folded conformations reached 0.78. During the final 50ns it also remained above 70%. This value is higher by approximately 10% than the results obtained by others [38] for REMD simulations with the Amber99sb force field and it is most likely caused by the different solvent model we used. Generally, we also obtained a higher native fraction at 280K compared to other versions of Amber force fields [39,41,42] used in simulations started from the crystallographic conformation.

The melting curve values computed from CABS-initiated simulations at higher temperatures for the first 50 ns and the last 50 ns also agree with each other. Differences are minimal, and we can assume that owing to the CABS models we obtain convergence very quickly. Final melting curves for both starting options were nearly identical (within the error limits) in either force field.

The analysis of energy distributions of neighboring replicas showed, that the application of the CABS-generated conformations does not change the replica exchange probability for either force field. For Amber99sb it is *ca.* 0.19 and 0.17 for OPLSAA. Overlapped potential energy histograms are presented in Figure 2 for the OPLSAA force field and the CABS starting option. Histograms are broader for higher temperature replicas and the exchange probability for them is a little higher.

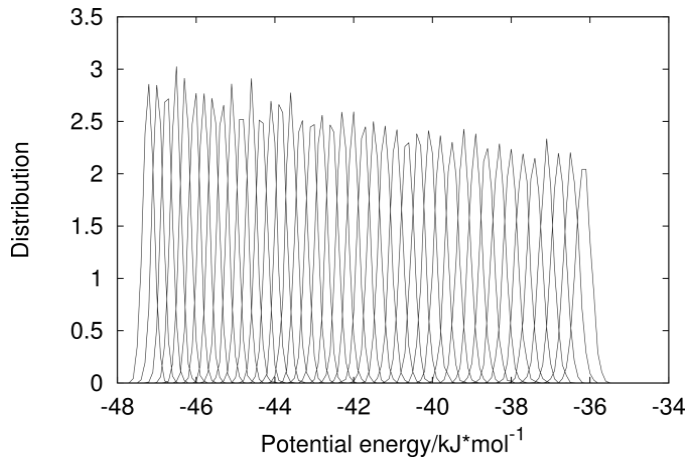

**Figure 2.** Potential energy distribution for each temperature replica divided by the number of particles in the system for the OPLSAA force field with CABS-generated starting conformations.

*2.2. Secondary Structure Propagation at 300 K*

With the CABS-starting option, native-like conformations are stable almost immediately in the beginning of simulation for the 300 K replica for both force fields (Figure 3). At this temperature there is domination of conformations with the β-hairpin structure featuring the correct pattern of native-like hydrogen bonds between the peptide backbone fragments. In the Amber99sb force field, conformations containing 4 (out of 5) main hydrogen bonds are the most stable in contrast to the OPLSAA force field for which structures with the complete pattern of main native-like hydrogen bonds are more common.

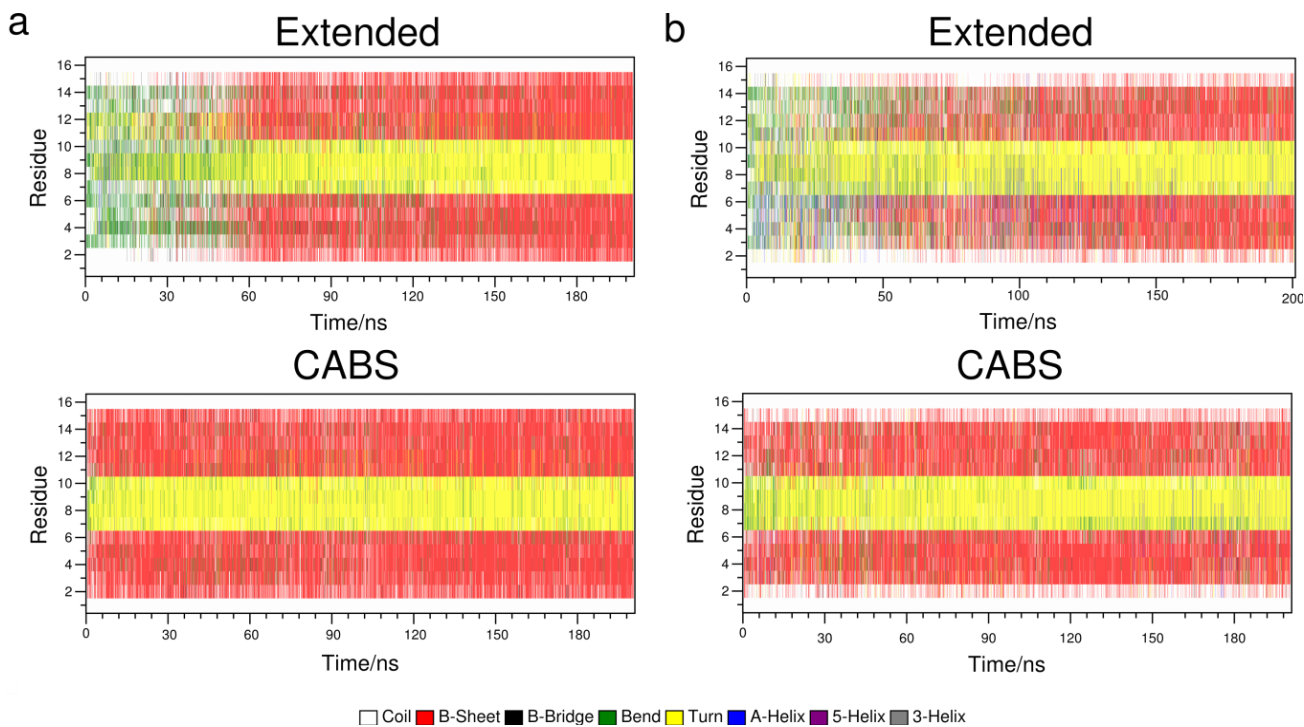

**Figure 3.** Plot of a secondary structure formed on every residue of the β-hairpin for two starting options: Extended conformations (upper panel) and CABS-generated conformations (lower panel) at 300 K. (**a**) diagrams for the OPLSAA force field; and (**b**) diagrams for the Amber99sb force field.

Generally, in each case the fraction of near-native β-hairpins remains constant at 300K after ca. 150ns, which can be another indicator of simulation convergence [56]. It holds at the level of ca. 85% for the OPLSAA force field and ca. 60% for Amber99sb (note that the stability of secondary structure motifs, as well as other features of folding dynamics, may be strongly force-field dependent [57]).

## 3. Methods

First, we performed a Monte Carlo (MC) isothermal simulation with the CABS [8] reduced model (available online [58]) starting from the extended conformation of β-hairpin from B1 domain of protein G(PDB code: 2GB1, residues 41 to 56). The reduced simulation temperature was near the transition point. In this way, we generated 10,000 snapshots. During the simulation, multiple transitions between near-native and fully unfolded ensembles were observed (Figure 4a). The high energy ensemble consists mainly of very loosely collapsed structures with the C-alpha Root Mean Square Deviation (CRMSD) near 4 Å. The low energy ensemble consists of conformations with the CRMSD oscillating around 2.5 Å. The transition to the native-like cluster is cooperative through a low-density region of states. Subsequently, we randomly selected 42 structures spanning the whole energy range. Most of them belonged to the most populated basin with a value of CABS energy around -90 (see Figure 4a), and therefore these were mainly selected for the REMD simulation. The selected conformations were subjected to a three-step reconstruction and minimization procedure [13,59] consisting of the following steps: (i) protein backbone reconstruction from the C-alpha trace [60], (ii) reconstruction of the side chains from the backbone chain [61] and (iii) a short minimization step, in vacuum, with frozen alpha carbons (example rebuilt models are presented in Fig. 4b). The applied reconstruction method was shown to produce physically sound models and a proper ranking of the quality of the models (distance from the native structure) when post-minimization all-atom energy was used as the ranking criterion [59]. According to all-atom energy, high energy structures were introduced to the high temperature replicas in the REMD simulation, with the low energy ones at low temperatures. Prior to the MD simulations, input structures were additionally minimized with the steepest descent method and equilibrated for 200 ps at a constant volume.

To test several variants of the method, four REMD simulations were performed with different starting conformations and different MD setups:

- Simulation #1 was conducted with 42 replicas (initialized with CABS output models as described above) with the replica exchange trial every 1 ps. Simulation time was 200 ns per replica. We used a dodecahedron simulation box containing 2087 water molecules and Na$^+$ and Cl$^-$ ions at a concentration of 0.15 M. Periodic boundary conditions were applied. Forty two temperature replicas were distributed in a range of 280 K–562 K. The OPLS-AA [18] force field was used with the spc [62] model for water. Bonds were constrained using the LINCS [63] algorithm.

  The equations of motion were integrated using a leap-frog algorithm[64] with a time step of 2fs. Non-bonded electrostatic interactions were computed using the particle-mesh Ewald [65] method and van der Waals interactions with a simple cut-off of 1 nm.
- Simulation #2 was conducted in the same way as #1 using the extended peptide as a starting conformation.

  The same procedures (#1 and #2) were carried out using AMBER99sb [66] (two simulations).
  The total simulation time was 33.6 μs.

We assigned conformations to the folded population according to the condition of Best and Mittal [38]. Briefly, conformations whose d$_{RMS}$ was less than 0.15Å were included in the near-native cluster. d$_{RMS}$ was calculated as follows:

$$d_{RMS} = \sqrt{\frac{\sum_{i,j}\left(r_{ij} - r_{ij}^0\right)^2}{N_{bb}}} \, , \, |i-j| > 2 \tag{1}$$

where $N_{bb}$—number of backbone native contacts; $r_{ij}$—distance between backbone atom $i$ and $j$; $r_{ij}^0$—distance between backbone atom $i$ and $j$ in the native conformation

In this equation we included only the backbone atoms (CA, C, N, O) which were at a maximum distance of 4.5Å in the native conformation.

All REMD simulations were done with the GROMACS package (Version 4.5.3) [67]. Analysis was performed with Bioshell [68,69] and dssp [70]. Visualization was prepared with the PyMOL Molecular Graphics System. [71]

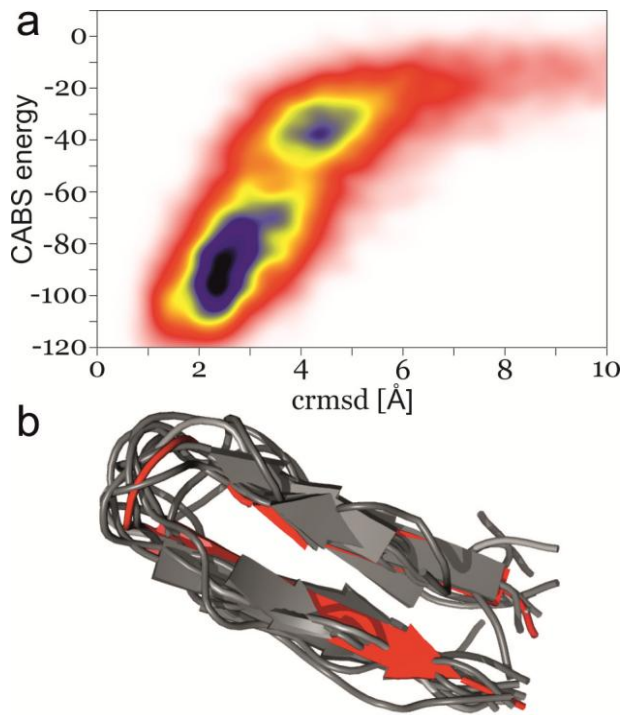

**Figure 4.** (**a**) Conformation density map as a function of CABS energy *vs.* C-alpha Root Mean Square Deviation (CRMSD); (**b**) example rebuilt conformations, extracted from the CABS trajectory. These models are selected so as to preserve a similar CRMSD to the native C-terminal β-hairpin of the 2GB1 protein, which is colored red.

## 4. Conclusions

Simulations of the GB1 β-hairpin in explicit solvent started from an extended conformation require many temperature replicas and long simulation times to equilibrate the system [38,56]. Our method helps to overcome these difficulties by merging CG modeling and all-atom REMD. We have found that by using CABS-generated conformations as the starting option for REMD, the time needed to reach the equilibrium state is dramatically reduced from hundreds to tens of nanoseconds. For Amber99sb and OPLSAA force fields, the convergence of average quantities derived from simulations occurs several times faster. This conclusion is based on the analysis of the number of folded replicas during the simulation and coincidence of melting curves depending on the chosen simulation period (Figure 1). For the majority of replicas we instantly obtained native-like conformation fraction values that were very similar to the experimental ones [50]. They were also in agreement with other REMD simulations in Amber99sb and OPLS-AA force fields [38]. Secondary structure propagation at 300K additionally reveals the effectiveness of our approach. The efficiency of simulations, both CG and all atom, may be improved by optimization of the temperature set for replica sampling [72]. Optimally allocated replicas facilitate fast flux of conformations through the temperature space. The major shortcoming of the method results from rather limited parameterization of CG force fields. For instance, the CG calculations cannot be conducted for proteins whose amino acid side chains interact with ligands or were chemically modified.

Importantly, the developed method has a potential application in protein structure refinement [73]. In our test we obtained improvement of β-hairpin native contacts and the secondary structure. Starting from the conformations with the average of 1.6 (of 5) native-like hydrogen bonds connecting the main chain, we obtained dominating clusters with representative conformations having four (for Amber99sb) and five (for OPLSAA) corresponding hydrogen bonds. To summarize, our investigation indicates that CABS-generated conformations are a great supplement to all-atom techniques, especially Replica-Exchange Molecular Dynamics. Our approach significantly reduces the REMD computation cost for peptides for which convergence time is a significant obstacle to obtain thermodynamic data [44,51,74,75]. It could be used for faster but still accurate studies of protein folding thermodynamics.

Since the CABS model allows for the prediction of folding dynamics of proteins much longer than the β-hairpin [10-14], the presented method offers the possibility of atomic-level characterization of larger protein systems than using REMD alone.

**Acknowledgements**

The authors acknowledge the support from the Foundation for Polish Science TEAM project (TEAM/2011-7/6) cofinanced by the European Regional Development Fund operated within the Innovative Economy Operational Program, and from the Polish National Science Center (Grant No. NN301071140). Calculations were performed in the Interdisciplinary Center for Mathematical and Computational Modeling (ICM) of the University of Warsaw (Grant G43-9).

The authors would like to thank Michal Jamroz for the great contribution to the preparation of REMD simulations.